\pdfoutput=1 
\documentclass[runningheads]{llncs}
\usepackage[T1]{fontenc}
\usepackage{graphicx}
\usepackage{amsmath,amsfonts,bm}

\def\eqref#1{equation~\ref{#1}}

\def\1{\bm{1}}

\DeclareMathAlphabet{\mathsfit}{\encodingdefault}{\sfdefault}{m}{sl}
\SetMathAlphabet{\mathsfit}{bold}{\encodingdefault}{\sfdefault}{bx}{n}

\begin{document}
\title{AutoMO-Mixer: An automated multi-objective Mixer model for balanced, safe and robust prediction in medicine}
%
%
\author{Xi Chen\inst{1} \and
Jiahuan Lv\inst{1} \and
Dehua Feng\inst{1} \and
Xuanqin Mou\inst{1} \and
Ling Bai\inst{1} \and
Shu Zhang\inst{1} \and
Zhiguo Zhou\inst{2} \thanks{corresponding author}}
\authorrunning{F. Author et al.}
%
\institute{School of Information and Communication Engineering, Xi'an Jiaotong University, Xi'an, China
\email{xi\_chen@mail.xjtu.edu.cn}\and
Department of Biostatistics and Data Science, University of Kansas Medical Center, Kansas City, KS\\
\email{zgzhou2013@gmail.com}}
\maketitle              
\begin{abstract}
Accurately identifying patient’s status through medical images plays an important role in diagnosis and treatment. Artificial intelligence (AI), especially the deep learning, has achieved great success in many fields. However, more reliable AI model is needed in image guided diagnosis and therapy. To achieve this goal, developing a balanced, safe and robust model with a unified framework is desirable. In this study, a new unified model termed as automated multi-objective Mixer (AutoMO-Mixer) model was developed, which utilized a recent developed multiple layer perceptron Mixer (MLP-Mixer) as base. To build a balanced model, sensitivity and specificity were considered as the objective functions simultaneously in training stage. Meanwhile, a new evidential reasoning based on entropy was developed to achieve a safe and robust model in testing stage. The experiment on an optical coherence tomography dataset demonstrated that AutoMO-Mixer can obtain safer, more balanced, and robust results compared with MLP-Mixer and other available models.

\end{abstract}
\section{Introduction}
With the development of modern medicine, medical image has become an essential tool to carry out personalized and accurate diagnosis. Due to the strong ability to analyze image, deep learning has been widely used in medical image analysis and has achieved great success~\cite{success1,success2} in the past years. However, many current available models can also lead to unreliable predictions. For example, the car's perception system misclassified the white part of the trailer into the sky, resulting in a fatal accident~\cite{car}. As such, different from other application fields such as face recognition, nature image classification, model reliability is more important in medicine as it is related to human life and health. On the other hand, we not only need to obtain the accurate prediction results, but also need to know whether the outcome is reliable or not. To realize this abstract goal by considering the clinical needs, we believe that building a unified model to achieve balance, safe and robust is desirable.

Currently, most prediction models use a single objective (e.g., accuracy, AUC)~\cite{mo1,mo2} function in the model training. However, the imbalanced sensitivity and specificity may result in higher rate of missed diagnosis~\cite{mo3}. Therefore, a multi-objective model which considers sensitivity and specificity simultaneously is needed. So far, there have been some studies on multi-objective optimization~\cite{mo4,mo5}.

Furthermore, since most models are data-driven based strategy, it is hard to evaluate whether the prediction outcome for an unseen sample is reliable or not. A possible solution is evaluating the model output by introducing a “third party” to independently estimate the model reliability or uncertainty. There have been several studies on uncertainty estimation for deep learning. ~\cite{safe1} proposed a framework based on test-time data augmentation to quantify the diagnostic uncertainty in deep neural networks. ~\cite{safe2} used the prediction of the augmented images to obtain entropy to estimate uncertainty.

Meanwhile, it is found that the model built based on the dataset collected from one institution always obtain bad performance when the testing dataset is from another institution~\cite{u1,u2,u3}, demonstrating the poor robustness. On the other hand, a reliable model should always work well across the multiple institutions. Several studies have investigated this issue. Adversarial attack is one of the most serious factors that cause models not to be robust~\cite{u4}. Some attackers perturbated test reports to obtain medical compensation~\cite{u5}. Adversarial examples lead to wrong decisions that can cause dangerous effects on the patient’s life ~\cite{u6}. ~\cite{u7,u8} evaluated the robustness of the model with adversarial attacks.

In summary, there have been several studies on building balanced, safe and robust model independently, but there is no unified framework that can achieve three goals simultaneously. As such, a new automated multi-objective Mixer (AutoMO-Mixer) model based on multiple layer perceptron Mixer (MLP-Mixer) is developed in this study to build a more reliable model. In AutoMO-Mixer, both sensitivity and specificity were considered as the objective functions simultaneously and a Pareto-optimal model set can be obtained through the multi-objective optimization~\cite{imia} in training stage. In testing stage, the Pareto-optimal models with balanced sensitivity and specificity were chosen so as to improve model balance. To obtain safer and more robust model, evidential reasoning based on entropy (ERE) approach was developed to fuse the outputs of Pareto-optimal models to obtain the final outcome. The experimental studies on optical coherence tomography (OCT) dataset demonstrated that AutoMO-Mixer can outperform MLP-Mixer and other deep learning models, and more balanced, robust and safer results can be achieved as well.

\section{Method}
\subsection{Overview}
\begin{figure}
\centering
\includegraphics[width=\textwidth]{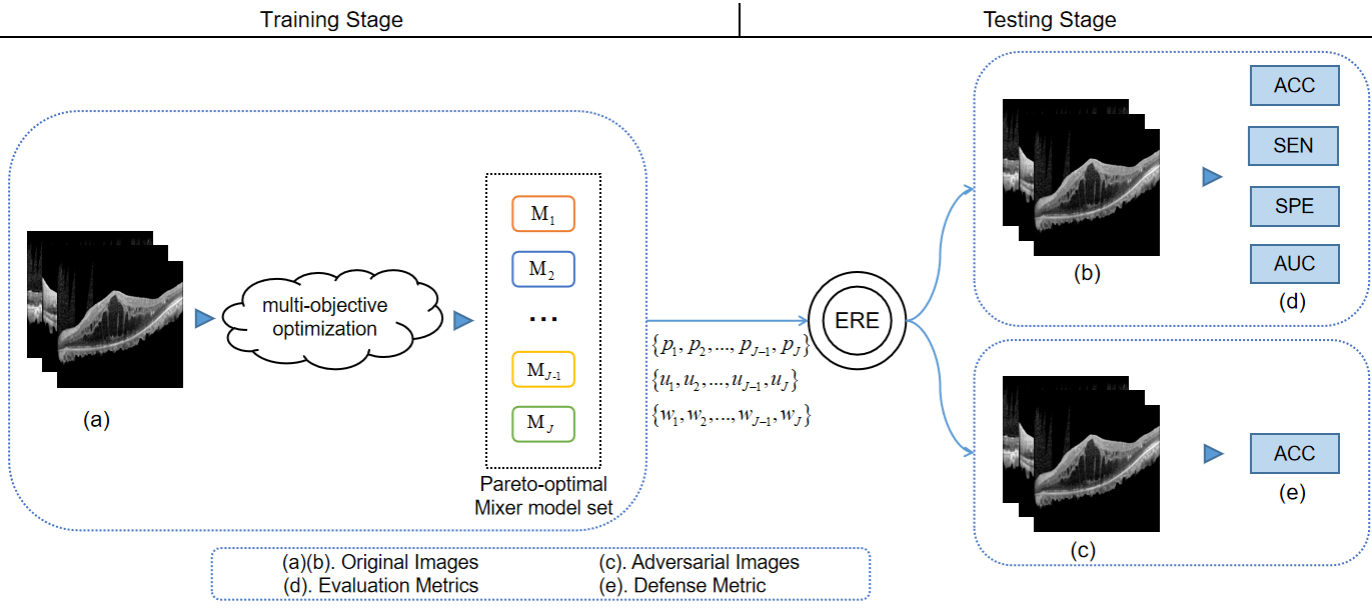}
\caption{The framework of AutoMO-Mixer model.} \label{AutoMO-Mixer}
\end{figure}

The framework of AutoMO-Mixer is shown in Fig.~\ref{AutoMO-Mixer}, which consists of training and testing stages. To build a balanced model, both sensitivity and specificity are considered as objective functions simultaneously in training stage, and a Pareto-optimal model set is generated then. To build a safer and more robust model, ERE strategy is developed to fuse the probability outputs of multiple Pareto-optimal models in testing stage.

\subsection{MLP-Mixer}
Since the computational complexity is increased sharply when there are more parameters in multi-objective learning, it is better to have fewer parameters in model training. The recently proposed MLP-Mixer~\cite{mixer} model is a full MLP architecture. Compared with CNN, the convolutional layer is removed from MLP-Mixer, leading to decreasing the scale of the architecture parameters sharply. On the other hand, MLP-Mixer can achieve similar performance to CNN~\cite{mixer}. Therefore, it is a better choice in multi-objective learning.

\subsection{Training stage}
In training stage, sensitivity denoted by ${f_{spe}}$ and specificity denoted by ${f_{sen}}$ are considered as objective functions simultaneously, they are:
\begin{equation}
  {f_{sen}} = \frac{{TP}}{{TP + FN}} \hfill \\
\end{equation}
\begin{equation}
  {f_{spe}} = \frac{{TN}}{{TN + FP}} \hfill \\
\end{equation}
where \emph{TP} and \emph{TN} represent the number of true positives and true negatives, \emph{FP} and \emph{FN} are the number of false positives and false negatives, respectively.

\begin{figure}[t]
\centering
\includegraphics[scale=0.6]{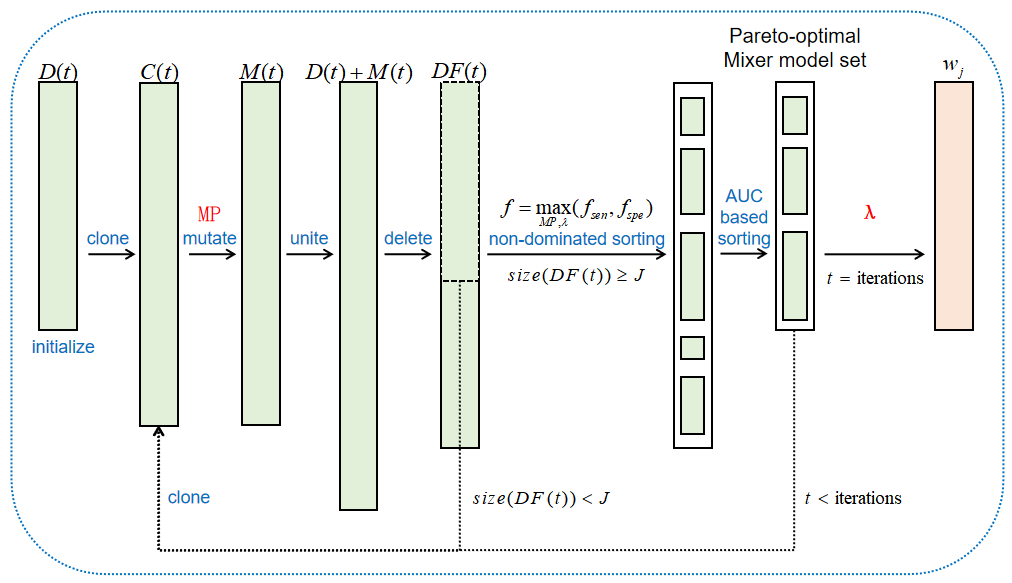}
\caption{The illustration of training stage.} \label{training stage}
\end{figure}

Assume ${\rm M} = \{ {{m}_1},...,{{m}_q}\} $ denotes the MLP-Mixer model, where q represents the number of model parameters. To obtain the balanced models, we aim to maximize ${f_{sen}},{f_{spe}}$ simultaneously, and an iterative multi-objective immune algorithm (IMIA)~\cite{imia} is used. IMIA consists of six steps: initialization, cloning, mutation, deletion, update, and termination. First, the initial model set denoted by $D(t) = \{ {{\rm M}_1},...,{{\rm M}_N}\} $ is generated , where ${{\rm M}_i} = \{ {m_{i1}},...,{m_{iq}}\} ,i = 1,2,...,N$. Then the models with higher ${f_{sen}},{f_{spe}}$ will be replicated using the proportional cloning method. In the third step, a probability of mutation is randomly generated for each model, and the model performs mutation when its probability is larger than the mutation probability (MP). After the mutation, the new models are generated. If some models have same sensitivity and specificity, only one model is remained. Then the model set size is kept through AUC based non-dominated sorting strategy. The training process will not stop until the maximum number of iterations is reached.  Finally, the Pareto-optimal Mixer model set is generated, where the model set size is J. Since the two hyperparameters MP and $\lambda $ may affect the model performance, Bayesian optimization~\cite{bys} is used to optimize the hyperparameters. The illustration of the training phase is shown in Fig.~\ref{training stage}.

\subsection{Testing stage}
\begin{figure}
\centering
\includegraphics[scale=0.49]{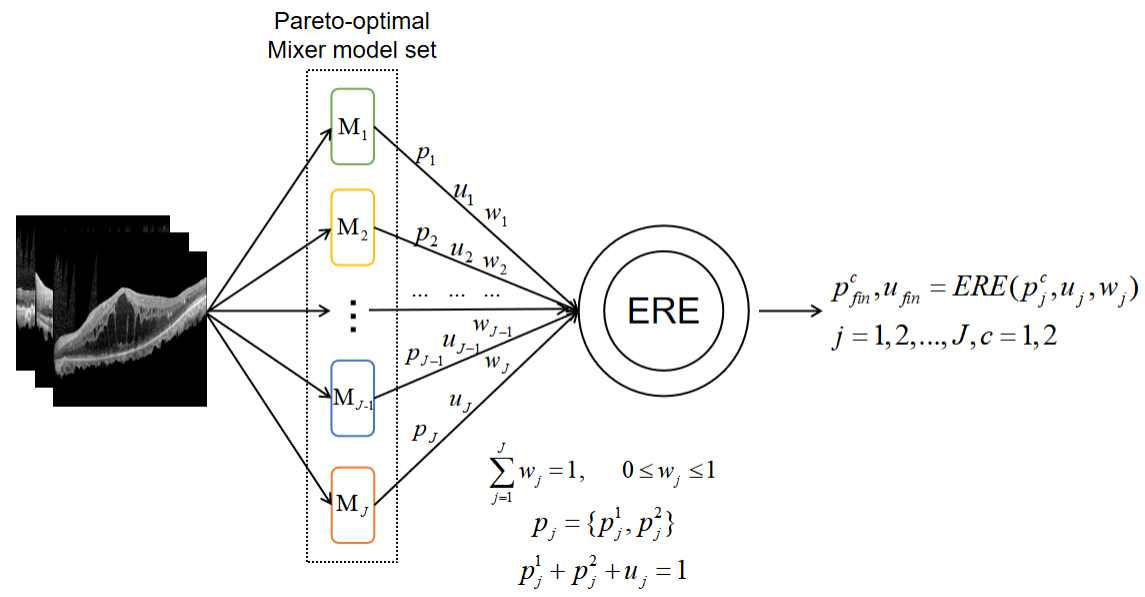}
\caption{The illustration of testing stage.} \label{testing stage}
\end{figure}

In testing stage, the probability outputs of Pareto-optimal models are fused through the evidential reasoning~\cite{er1,er2} based on entropy approach. The workflow is shown in Fig.~\ref{testing stage}.

\subsubsection{Weight calculation} Since the performance of different Pareto-optimal models cannot be the same, the weight for each model should be estimated, which is denoted by ${w_j}$. As the balanced model between sensitivity and specificity is desired, the ratio between them is considered in the weight calculation, that is  $\frac{{{f_{sen}}}}{{{f_{spe}}}}$ or $\frac{{{f_{spe}}}}{{{f_{sen}}}}$. When the ratio is less than 0.5 or greater than 1, the model is considered as extreme imbalance, setting ${w_j}$ as 0. Meanwhile, AUC is a good measure for model reliability, it is also considered. The expression of ${w_j}$ is as follows:
\begin{equation}
{w_j} = 
\begin{cases}
\lambda \frac{{f_{sen}^j}}{{f_{spe}^j}} + (1 - \lambda )AU{C_j},\;\;\;\;when\;0.5 \leq \frac{{f_{sen}^j}}{{f_{spe}^j}} \leq 1 \hfill \\
\lambda \frac{{f_{spe}^j}}{{f_{sen}^j}} + (1 - \lambda )AU{C_j},\;\;\;\;when\;0.5 \leq \frac{{f_{spe}^j}}{{f_{sen}^j}} \leq 1,j\; = \;1,2,...,J \\
0\;\;\;\;\;\;\;\;\;\;\;\;\;\;\;\;\;\;\;\;\;\;\;\;\;\;\;\;\;\;\;\;\;\;\;\;Other\;situations \\
\end{cases}
\end{equation}
where $\lambda $ indicates the importance of balance, and $1-\lambda $ indicates the importance of AUC. After calculating the ${w_j}$ for each model, the weights are normalized.

\subsubsection{Uncertainty estimation} Test-time data augmentation (TTA)~\cite{safe1,safe2} is used to perform useful estimates of model uncertainty. The test image is fed into model ${M_j},j = 1,2,...,J$ to generate the probability output $p_{_j}^c,c = 1,2$, where $p_j^1 + p_j^2{\text{ = }}1$. The original test image is enhanced T times to generate prediction $p_{j,t}^c,t = 1,2,...,T$. The mean class probability $\overline {p{}_j^c} $ and the uncertainty ${u_j}$ are:
\begin{equation}
\overline {p_{_j}^c}  = \frac{1}{T}\sum\limits_{t = 1}^T {p_{_{j,t}}^c} \begin{array}{*{20}{c}}
  ,&{c = 1,2} 
\end{array}
\end{equation}
\begin{equation}
{u_j} =  - \sum\limits_{c = 1}^2 {\overline {p_j^c} } \log (\overline {p_j^c} )
\end{equation}

To satisfy the conditions of the ERE strategy, $p_j^c$ and ${u_j}$ are normalized so that $p_{_j}^1 + p_{_j}^2 + {u_j} = 1$.

\subsubsection{ERE strategy}Assume that the output probability for each model is denoted by ${p_j} = \{ p_j^1,p_j^2\} , p_{{j}}^1+p_{{j}}^2 \leq 1,j = 1,2,...,J$. If $p_{{j}}^1+p_{{j}}^2< 1$, it shows that the jth model has uncertainty ${u_j}$ on its output. Then the final output probability $p_{{fin}}^c,c=1,2$ and uncertainty ${u_{fin}}$ are obtained through the ERE fusion strategy. that is:
\begin{equation}
  p_{fin}^c,u_{fin} = ERE(p_j^c,{u_j},{w_j}),j = 1,2,...,J,c = 1,2
\end{equation}
where ERE is:
\begin{equation}
  p_{fin}^c = \frac{{\mu  \times [\prod\limits_{j = 1}^J {({w_j}p_j^c + 1 - {w_j}(p_{{j}}^1+p_{{j}}^2))}  - \prod\limits_{j = 1}^J {(1 - {w_j}(p_{{j}}^1+p_{{j}}^2))} ]}}{{1 - \mu  \times [\prod\limits_{j = 1}^J {(1 - {w_j})} ]}},c = 1,2
\end{equation}
\begin{equation}
{u_{fin}} = \frac{{\mu  \times [\prod\limits_{j = 1}^J {(1 - {w_j}(p_j^1 + p_j^2))}  - \prod\limits_{j = 1}^J {(1 - {w_j})} ]}}{{1 - \mu  \times [\prod\limits_{j = 1}^J {(1 - {w_j})} ]}}
\end{equation}
The normalized factor $\mu $ is:
\begin{equation}
  \mu  = {[\sum\limits_{c = 1}^2 {\prod\limits_{j = 1}^J {({w_j}p_j^c + 1 - {w_j}(p_{{j}}^1+p_{{j}}^2))}  - \prod\limits_{j = 1}^J {(1 - {w_j}(p_{{j}}^1+p_{{j}}^2))} } ]^{ - 1}}
\end{equation}

\subsection{Robustness evaluation}
In this study, fast gradient sign method (FGSM)~\cite{fgsm} is used to disturb the original samples, which is a white box attack with full information of the models. Adversarial samples are generated by the following formula:
\begin{equation}
{x^a} = x + \delta
\end{equation}
where ${x^a}$ represents the adversarial sample, $x$ represents the original sample. $\delta$ represents the perturbation. The degree of perturbation is controlled by $\varepsilon $. In our study, ACC is used to evaluate robustness~\cite{u8}.

\section{Experiments}
\subsection{Experimental setup}
The dataset used in this study was collected from the Second Affiliated Hospital of Xi'an Jiaotong University (Xi’an, China), including 228 patients with Choroidal neovascularization (CNV) and cystoid macular edema (CME) between October 2017 and October 2019. First, OCT images of each patient were acquired via the Heideberg Retina Tomograph-IV (Heidelberg Engineering, Heidelberg, Germany). These patients were then injected with anti-vascular endothelial growth factor (anti-VEGF) and the evaluations were made after 21 days. Among them, anti-VEGF was effective for 171 patients, and the remaining 57 patients had no sign of effectiveness. The study was approved by the Research Ethics Committee, and each patient provided written informed consent. In our study, we built a binary classifier to determine whether anti-VEGF would be effective for patients using OCT images. In the training stage, there were 135 effective cases and 44 ineffective cases. In the testing stage, there were 34 and 12, respectively, in these two classes.

Before being fed into the model, all the images were resized into 224 x 224. MP and $\lambda $ were set to 0.5 and 0.8, respectively. The MLP-Mixer contains four parameters, these settings are shown in Table~\ref{parameters}.

\begin{table}[t]
\centering
\caption{The range of values for MLP-Mixer network structure parameters.}\label{parameters}
\begin{tabular}{cc} 
\hline 
\bfseries parameters & \bfseries range of values \\ 
\hline
\ Number of layers & [2, 3, 4] \\
\hline
\ Hidden size C & 256*[1, 1.2, 1.4, 1.6] \\
\hline 
\ MLP dimension Ds & 196*[2, 3, 4, 5] \\
\hline
\ MLP dimension Dc & 256*[2, 4, 6, 8, 10] \\
\hline
\end{tabular}
\end{table}

As AutoMO-Mixer was built based on MLP-Mixer and ResNet-18 is a classical deep learning model, they were used in comparative study. The four parameters in MLP-Mixer network were set to 5, 256, 392, 1024, respectively, and transfer learning was used on ResNet-18 as pre-training. Sensitivity (SEN), specificity (SPE), area Under Curve (AUC), and accuracy (ACC) were used for evaluation. All the experiments were performed five times, and mean and standard deviation were evaluated.

\subsection{Results}
The evaluation results on MLP-Mixer, ResNet-18 and AutoMO-Mixer are shown in Table~\ref{evaluation}. In this study,  $\frac{{\min (SEN,SPE)}}{{\max (SEN,SPE)}}$ was used to assess the balance of the model. It can be seen that AutoMO-Mixer model is the most balanced. In addition, both the AUC and ACC of the AutoMO-Mixer are better than the other two models.

\begin{table}[t]
\centering
\caption{The evaluation results on OCT dataset.}\label{evaluation}
\begin{tabular}{cccccc} 
\hline 
\bfseries models & \bfseries SEN & \bfseries SPE & \bfseries AUC & \bfseries ACC & \bfseries $\frac{{\min (SEN,SPE)}}{{\max (SEN,SPE)}}$ \\
\hline
\ MLP-Mixer & 0.611±0.052 & 0.703±0.077 & 0.709±0.041 & 0.671±0.038 & 0.869 \\
\hline
\ ResNet-18 & 0.728±0.075 & 0.706±0.071 & 0.791±0.046 & 0.714±0.052 & 0.970 \\
\hline
\bfseries AutoMO-Mixer & 0.778±0.000 & 0.779±0.000 & 0.844±0.000 &  0.779±0.000 & \bfseries 0.999 \\
\hline
\end{tabular}
\end{table}

\subsubsection{Safety evaluation} In this study, the uncertainty estimation was used to measure model safety. If the performance of the model can improve as the number of test samples with high uncertainty decreases, it is indicated that the model is safe. The entire test samples were arranged from smallest to largest in order of uncertainty, with the maximum uncertainty being 0.4245, the upper quartile being 0.4206, the median being 0.4165, and the lower quartile being 0.4045. Samples with less uncertainty than them were grouped into four cohorts, and the evaluation results are shown in Table~\ref{uncertainty}. It can be seen that the lower the cutoff uncertainty is, the better the model's performance is, indicating our model can assess whether the prediction is safe based on uncertainty.

\begin{table}[t]
\centering
\caption{Model performance of the test cohorts stratified by the uncertainty.}\label{uncertainty}
\begin{tabular}{ccccc} 
\hline 
\bfseries uncertainty & \bfseries SEN & \bfseries SPE & \bfseries AUC & \bfseries ACC \\
\hline
\ 0.4245 & 0.778 & 0.779 & 0.844 & 0.779 \\
\hline
\ 0.4206 & 0.783 & 0.796 & 0.860 & 0.792 \\
\hline
\ 0.4165 & 0.818 & 0.829 & 0.823 & 0.827 \\
\hline
\ 0.4045 & 1.000 & 0.895 & 1.000 & 0.920 \\
\hline
\end{tabular}
\end{table}

\subsubsection{Robustness} 
\begin{figure}[t]
\centering
\includegraphics[scale=0.55]{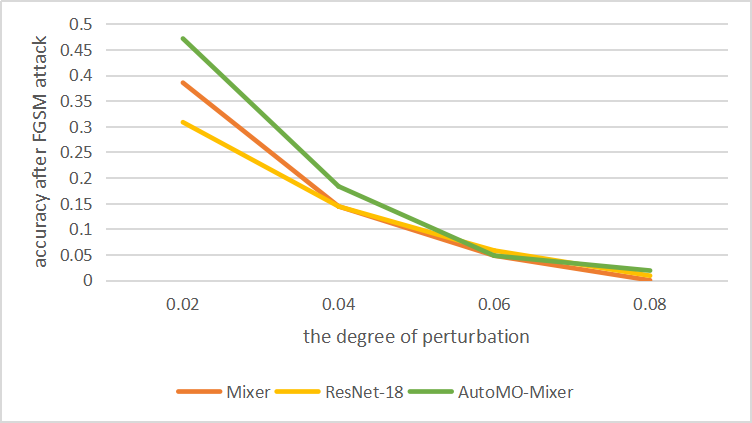}
\caption{Comparison of the robustness between the AutoMO-Mixer, ResNet-18 and AutoMO-Mixer models.} \label{robustness}
\end{figure}

After the original samples were attacked by FGSM, indistinguishable adversarial samples were generated. We measured the accuracy of adversarial samples in each model in Fig.~\ref{robustness}. It is obvious that except slightly less when $\varepsilon $=0.06, the robustness of AutoMO-Mixer is better than the other as a whole.

\section{Conclusions}
In this study, a new model termed as AutoMO-Mixer was developed for image guided diagnosis and therapy. In AutoMO-Mixer, sensitivity and specificity were considered as the objective functions simultaneously and a Pareto-optimal Mixer model set can be obtained in training stage. In testing stage, ERE was used to obtain safer and more robust results. The experimental results on OCT dataset showed that AutoMO-Mixer outperformed MLP-Mixer and ResNet-18 in balance, safe and robustness.

%
%
%
%

\end{document}